# Concurrent risks of dam failure due to internal degradation, strong winds, snow and drought

Chris Collier*, Alan Gadian, Ralph Burton and James Groves

National Centre for Atmospheric Science, School of Earth & Environment, University of Leeds, Leeds, LS2 9JT, Yorkshire, United Kingdom; * corresponding author <chris.collier@ncas.ac.uk>

**Abstract**

The chance (or 'probability') of a dam failure can change for various reasons such as structural degradation, the impacts of climate change and land-use change. Similarly the consequences of dam failure (flooding) can change for many reasons such as growth in the population in areas below a dam. Consequently both the chance that a dam might fail and the likely consequences of that failure can change over time. It is therefore crucial that reservoir safety risk analysis methods and decision-making processes are able to support (as a minimum) 'what-if' testing (or 'sensitivity testing') to take into account these changes over time to gauge their effect on the estimated risk of dam failure. The consequences of a dam failure relate to the vulnerability and exposure of the receptors (for example, people, property and environment) to floodwater. Also the probability of dam failure varies with age, design and construction of the dam.

Spillway failure may be caused by the dissipation of energy from water flowing down the spillway, and embankment erosion (scour) may be caused by a dam overtopping. The occurrence of these events depends upon the dam design and the likelihood of extreme rainfall, also in the case of overtopping wind-driven waves on the reservoir surface. In this study the meteorological situations of notable recent events i.e. the Boltby, North Yorkshire incident, 19 June 2005 in which the dam almost overtopped, and the spillway failure of the Ulley Dam near Rotherham at the end of June 2007, are studied. The WRF numerical model will be used to indicate how these meteorological situations might be maximized, and be coupled with the occurrence of other failure modes such as the likelihood of internal dam failure assessed from previous work by government panel engineers.

**1. Introduction**

The chance of a flood occurring downstream of the reservoir is based on consideration of the source of the threat (for example, an extreme rainfall event, internal erosion of the dam or earthquake), the performance or 'response' of the reservoir and dam when that threat occurs, and the nature of the downstream valley (the so-called pathway of risk). The consequences of a dam failure relate to the vulnerability and exposure of the receptors (for example, people, property and environment) to floodwater. Also the probability of dam failure varies with age, design and construction of the dam. There have been at least 41 safety problems at dams in England and Wales over the past six years (to December 2013), including 14 serious incidents.

Although there have been no failures resulting in loss of life in the UK since 1925, it has been suggested by Atkins (2013) that the UK has been fortunate and lucky. There have been many failures of what are known as 'small raised reservoirs' (SRRs) (up to a volume of 25,000 m$^3$) and many incidents. Flood events, particularly in 2007, showed that many SRRs were not maintained properly, or operated properly and that the risks posed by them were significant. This was recognised by Pitt (2008) who recommended a move to a risk based system and a move to include SRRs. Large reservoirs (reservoirs greater than 25,000 m$^3$) were required to be registered under the UK Reservoirs Act 1975, whereas the Flood and Water Management Act 2010 changed this such that a reservoir will now be considered: "large" if it is capable of holding 10,000 m$^3$ (as opposed to 25,000 m$^3$) of water or more; and "raised" if it is capable of holding water above the natural level of any part of the surrounding land.

The modes of dam failure include surface and internal erosion, crest fissuring, spillway breakup, a blocked overflow, collapse of outlet pipes and controlled holding of water at a low level. These problems may be related to the structure of the dam for example whether it is clay filled or not, subjected to long-term drought conditions, or extensive freezing and thawing. Spillway failure may be caused by the dissipation of energy from water flowing down the spillway, and embankment erosion (scour) may be caused by a dam overtopping. The occurrence of these events depends upon the dam design and the likelihood of extreme rainfall, also in the case of overtopping wind-driven waves on the reservoir surface. It is crucial that reservoir safety risk analysis methods and decision-making processes are able to support (as a minimum) 'what-if' testing (or 'sensitivity testing') to take into account these changes over time to gauge their effect on the estimated risk of dam failure.

A dam failure can result in local loss of life or damage to property. The consequences are not likely to be on a national scale, although the consequences can occur on a regional scale. A recent example is the Boltby, North Yorkshire incident, 19 June 2005 in which the dam almost overtopped. The dam was built in the 1880's and is located 1.5 km upstream of the village of Boltby. It is an approximately 150m long and 20m high earth-fill. The spillway consisted of a weir on the east side of the dam which leads to a masonry spillway channel. Since it is situated upstream of Boltby it is designated a category A dam (see later), such that the spillway must safely pass the Probable Maximum Flood (PMF) criterion. A raingauge reading taken at Hawnby, about 5km to the ENE of the dam, gave 64mm occurring in just over an hour with 125 mm falling in three hours, but the reading at Thirlby was lost as the gauge was overwhelmed. In this incident a temporary camp site at Duncombe Park near Helmsley in North Yorkshire was evacuated (Wass et al., 2008). Had the storm occurred 24 hours earlier and the dam failed completely 10,000 people sheltering in tents attending a rally would have occupied the site and been at risk. The flood occurred on rivers which had not experienced serious flooding in living memory.

An earthquake may cause a dam failure either directly, or through a concommitant landslip into the reservoir causing a large wave which overtops the dam. Such events have not occurred in the UK to date, but have occurred elsewhere. On October 9th, 1963 when the reservoir behind the Vajont Dam in Italy was being filled a section of rock weighing 150 million tonnes fell from the face of the nearby Monte Toc into the water. A huge swell occurred washing away several towns above the dam, while a second wave overtopped the Dam into the valley below, killing around 2,000 people and everything in its path.

In this paper we discuss the concurrent risks of a dam failure from a statistical point of view. However, this analysis is dependent upon knowledge of the probability of failure due to the likelihood that specific conditions occur. Of particular importance is rainfall approaching the design Probable Maximum Precipitation (PMP). We examine this for one case using an atmospheric numerical model in order to indicate an approach to estimating the statistical likelihood of occurrence of such extreme rainfall.

**2. Classification of dams in the UK**

Reservoir owners, operators and uses are responsible for making sure that their reservoirs operate safely and are properly managed. Reservoirs are classified on a consequence of failure basis as shown in Table 1. Figure 1a shows the distribution of Category A and B in England and Wales, and Figure 1b is an example from the Environment Agency Web site of the area likely to be flooded should dams fail near Pontefract, Yorkshire. Similar maps may be produced throughout England and Wales.

However, the high water level downstream of the Ulley reservoir located in Yorkshire during the event of 24-25 June 2007, and the threat of the dam failure, resulted in the closure of the M1 motorway for 40 hours, and the evacuation of 1,000 residents from the villages of Catcliffe, Whiston and Treeton. The Ulley dam was previously classified as Category C, but this incident suggests that the classification scheme needed to be reviewed. Similar occurrences support the need to carry out similar re-classification elsewhere.

The probability of a dam failure can change for various reasons such as structural degradation, the impacts of climate change and land-use change. Woldemichael et al. (2012) discuss the changes that dam reservoirs may make to the occurrence of extreme precipitation. Similarly the consequences of dam failure (flooding) can change for many reasons such as growth in the population in areas below a dam, or perhaps because the value of property below the dam has increased. Consequently both the chance that a dam might fail and the likely consequences of that failure can change over time.

Given that individual dam failures are independent events, it is possible that more than one failure could happen around the same time. If there are several dams built along the same river valley e.g. the Ladybower dams in the Peak District, it is conceivable that one failure may lead to another failure. In this case each failure becomes a dependent event.

### 3. The statistics of concurrent risks

Floods have the same probability of occurrence each year. A risk-based approach has largely replaced the traditional and potentially misleading term 'return period' as shown in Table 2.

In considering concurrent risks there are various types of events that need to be defined. We will look at each type following Chatfield (1983), considering specific events which are specified as $E_1$, $E_2$, $E_3$ etc.

1. **Mutually exclusive**

If two events $E_1$ and $E_2$ are mutually exclusive i.e. they cannot both occur, the probability that one of the mutually exclusive events occurs is the sum of their respective probabilities:

$$P(E_1 + E_2) = P(E_1) + P(E_2)$$

Here $P(E_1 + E_2)$ means that at least one of the events occurs. This could refer to internal dam erosion or very heavy rainfall.

2. **Not mutually exclusive**

Two events that are not mutually exclusive contain one or more sample points. Consider that internal dam erosion and very heavy rainfall do occur at the same time then,

$$P(E_1 + E_2) = P(E_1) + P(E_2) - P(E_1 E_2)$$

Here $P(E_1 E_2)$ is the probability that both $E_1$ and $E_2$ occur at the same time.

3. **Conditional probability**

Consider the probability of an event $E_1$ when it is not known that $E_2$ has occurred. In this case the conditional probability of $E_1$ given that $E_2$ has occurred is

$$P(E_1/E_2) = P(E_1 E_2) / P(E_2)$$

4. **Independent and dependent**

Two events $E_1$ and $E_2$ are said to be independent if $P(E_1) = P(E_1|E_2)$. When they are dependent then $P(E_1) \neq P(E_1|E_2)$.

5. **Joint events**

The probability of a joint event $(E_1 E_2)$ is given by

$$P(E_1|E_2) = P(E_1 E_2) / P(E_2) = P(E_2) P(E_1|E_2) = P(E_1) P(E_2|E_1)$$

These relations apply to both dependent and independent events. However, if the events are independent then $P(E_1|E_2) = P(E_1)$ and $P(E_1 E_2) = P(E_1) P(E_2)$

If there are three events then for example

$$P(E_1 + E_2 + E_3) = P(E_1) + P(E_2) + P(E_3) - P(E_1 E_2) - P(E_1 E_3) - P(E_2 E_3) + P(E_1 E_2 E_3)$$

And the other equations above may be extended similarly.

In assessing overall probability of failure the extent to which failure modes are dependent or independent of one another needs to be considered to apply the correct approach from the above. Events are dependent if the outcome of one event affects the outcome of another. For example, if one draws two coloured balls from a bag and the first ball is not replaced before you draw the second ball then the outcome of the second draw will be affected by the outcome of the first draw.

Where dependency is ignored the overall probability may be too high (although for dams the probabilities of failure are normally so small that this makes no practical difference to the outcome). Only the highest probability from each external threat may be considered, but for internal threats (from within the dam itself) all failure modes are summed as suggested by Defra/Ea (2013) and as shown in Table 3.

In cases where several dams retain the same reservoir, but would breach into the same valley, the same approach should be taken in combining the likelihood of failure of the dams, in that only the failure mode with the highest probability for all dams from one external threat is included, but that all internal threats are included.

The internal threats listed in Table 3 arise from structural degradation over time which may be enhanced by extended periods of drought. A meteorological drought is usually defined as an extended period of weather (usually around 3 weeks) where less than a third of the usual precipitation falls. Droughts are a relatively common feature of the weather in the United Kingdom, with one around every 5–10 years on average. These droughts are usually confined to summer, when a blocking high pressure system causes hot, dry weather for an extended period.

A hydrological drought can occur, after a relatively dry winter whereby the soil moisture storage, reservoirs and water table have not risen sufficiently to counteract the warm summer weather. These sort of conditions can go on over several years, even with above average rainfall at the time as the rainfall only slowly percolates through the water stores and replenishes them. This may lead to degradation within earth core dams. For example, the 16-month England and Wales rainfall total from May 1975 to August 1976 ranks the driest in the near 250-year England and Wales precipitation series for any 16-month period by a considerable margin (Rodda and Marsh, 2011). The 2010-2012 drought in England and Wales was similarly extreme (Kendon et al.(2013).

Crest and chute overtopping arise from reservoir levels being increased by extreme rainfall, and the occurrence of strong surface winds blowing at right angles to the dam. During the 1987 storm over England and Wales a maximum gust of 115 mph was observed at Shoreham-by-sea, West Sussex with sustained average wind speeds of 50 mph. The maximum observed gust speed at a low level site in Scotland at Fraserburgh is 142 mph.

The melting of lying snow has the potential to add to a flood produced by heavy rainfall, and hence increase the possibility of dam failure. The winter of 1946–1947 was one of the most significant winters to cause widespread flooding in the United Kingdom enhanced by a considerable snowmelt. Mid-March brought milder air to the country which thawed the snow lying on the ground. This snowmelt ran off the frozen ground straight into rivers and caused the widespread flooding. The weather factors which cause melting are temperature, wind speed and humidity.

Hough and Holliis (1997) estimated the annual maximum snow melt for periods of 3 to 168 hours. In addition they estimated the snow melt with return periods of 5, 25 and 50 years for these durations, and for a snow melt of 42 mm/day which was recommended for use in reservoir safety design studies. Table 4 shows estimates of return period and annual frequency for 24 hours snow melt equal to 42 mm at various stations throughout the United Kingdom. Note that for some high level stations annual frequencies are quite large e.g. for Malham Tarn it is 0.25 and Aviemore it is 0.125, whereas for low level stations in England and Wales it is very low $< 0.0001$.

Consider now the following two not mutually exclusive events:
$E1 = 5.10^{-5}$ a near PMP (see Collier et al., 2010)
$E2 = 6.10^{-4}$ internal threat
$P(E1E2) = 1.8.10^{-4}$ from Table 2
Hence using the equation above for $P(E1 + E2)$ then $P = 5.10^{-5} + 6.10^{-4} - 1.8.10^{-4} = 4.7.10^{-4}$. This value for the overall likelihood of failure is about two times larger than the overall probability given in Table 3.

Consider a near PMP rainfall event occurring at the same time as a snow melt of 42 mm/day, then at a high level station such as Malham Tarn the annual probability of failure for a joint event would be $P(E1|E2) = P(E1E2) / P(E2)$ where $E1 = 10^{-4} / 0.25 = 4.10^{-4}$. This is about twice the probability of the overall probability of failure given in Table 3.

**4. Numerical modelling**

Although the statistical approach does provide a useful insight into the probability of a dam failure, the use of an atmospheric numerical model can provide information on the probability of how physical processes might interact leading to dangerous conditions. We describe next the use of such models.

Collier and Hardaker (1996) describe a method of estimating PMP based upon a simple storm model. The model uses a parameter storm efficiency (E), which is the ratio of the total rainfall at the ground (cell-relative) to the amount of precipitable water (total cloud water condensed) in the representative air column during the storm. It is assumed that the air mass is saturated and the vertical humidity profile is represented by the dew-point temperature at the surface following the saturated pseudo-adiabatic lapse rate. The UK Flood Studies Reprt (FSA) describes the process of determining the storm efficiency. One problem with estimating E is that a good estimate of the dew-point temperature is needed. A more comprehensive numerical model is required than are represented in the simple model in order to investigate the impact of low-level convergence and increased understanding of storm microphysics and dynamics. This should lead to the definition of which factors are more important in producing heavy rainfall, and therefore improvements in techniques of estimating PMP.

The Weather Research and Forecasting (WRF) Model is a next-generation mesoscale numerical weather prediction system designed to serve both atmospheric research and operational forecasting needs. It features two dynamical cores, a data assimilation system, and a software architecture allowing for parallel computation and system extensibility. The model serves a wide range of meteorological applications across scales ranging from meters to thousands of kilometres. The effort to develop WRF began in the latter part of the 1990's and was a collaborative partnership principally among the National Center for Atmospheric Research (NCAR), the National Oceanic and Atmospheric Administration (represented by the National Centers for Environmental Prediction (NCEP) and the (then) Forecast Systems Laboratory (FSL)), the Air Force Weather Agency (AFWA), the Naval Research Laboratory, the University of Oklahoma, and the Federal Aviation Administration (FAA).

WRF allows researchers the ability to produce simulations reflecting either real data or observations, analyses) or idealized atmospheric conditions. The model physics is numerical code describing those processes (Table 5) not explicitly included in the basic dynamical and thermodynamical equations describing the earth's atmosphere. These processes are either too complicated to be explicitly included in the model based on their most fundamental physics laws (e.g. radiation and microphysics), or finer in scale than can be adequately represented by realizable grid resolutions (sub-grid scale turbulence, PBL transport). Yet, their effects on the resolvable scale flows and on the sensible weather (e.g., precipitation amount) have to be properly included for a model to accurately predict atmospheric behaviour for NWP purposes. Simplifications are typically made and variables (parameters) on the resolvable scales are often used in treating these processes; the resulting schemes are usually referred to as physics parameterizations.

To investigate the rainfall, which occurred on the $19^{th}$ June 2005 in North Yorkshire, three nested WRF numerical model domains over Northern England were used as follows:

D03 dx = 300 m 240 km square (about 800 x 800 grid points)
D02 dx = 1500 m 750 km square (about 500 x 500 grid points)
D01 dx = 7500 m 3750 km square (about 500 x 500 grid points)

At least 101 levels in the vertical were used. Convective parameterization were turned off except for the outer domain (D01). A series of experiments have been carried out to investigate the impact of changes to the model microphysical schemes, low-level convergence and orographic resolution. The microphysical schemes represent cloud and precipitation processes, and therefore of direct relevance to PMP. Further details of the microphysics schemes are given in Skamarock et al. (2008). Table 6 shows the different microphysical schemes tested. It was found that the WDM6 scheme produced the best results. In addition, planetary boundary layer wind convergence induced by atmospheric system dynamics and local orography is also of central importance. Comparing the rainfall generated by these processes with what actually occurred, it was found that about 50% of the rain resulted from the optimum choice of microphysical scheme, whereas 25% arose from using high resolution orography (50 m).

In Fig 3(b), neither the Thompson nor the Millbrandt scheme produced rainfall in this catchment region (cf. Figure 2). The Lin scheme produced a modest amount of rain (10mm); the Morrison scheme produced 30mm of rain and the WDM6 scheme exceeded 40mm. Note that the Morrison and WDM6 schemes forecast the location of the rainfall maximum very well (the latter scheme in particular), although the predicted amounts are far less those observed.

This shows (i) the sensitivity of NWP models to choice of microphysics scheme and (ii) that with a carefully chosen scheme the NWP model can, in fact, predict the correct location of intense rainfall. This suggests that a probability-based model using the information on failure modes discussed in section 3 could embrace the WRF model to assess the probability of a near PMP event.

## 5. Conclusion

Further work needs to be undertaken here, but it would appear that concurrent events can significantly increase the chances of failure. A probability-based model would appear to the way forward to assess the likelihood of current dam failure modes.

developments in the management of existing dams in the UK. A report was issued by Defra and the Environment Agency in collaboration with Ove Arup and Partners Ltd. in August 2011 as follows:

**Report:** SC080048/R1

**Title:** Modes of dam failure and monitoring and measuring techniques.

The most recent study of risk assessment for reservoir safety management was a guide produced by a consortium of experts let by HR Wallingford working with the Environment Agency and Atkins Ltd, Sayers & Partners, RAC Engineers & Economists and Samui Ltd., dated May 2013:

**Report:** SC090001/R1

**Title:** Guide to risk assessment for reservoir safety management. Volume 1: Guide

**Report:** SC090001/R2

**Title:** Guide to risk assessment for reservoir safety management. Volume 2: Methodology and supporting information

# List of tables

**Table 1**

| Dam category | Potential consequence of reservoir failure |
|---|---|
| A | At least 10 lives at risk and extensive property damage |
| B | Fewer than 10 lives at risk or extensive property damage |
| C | Negligible risk to human life, but some property damage |
| D | Negligible risk to human life and very limited property damage |

**Table 2**

| Annual probability (%) | Annual chance |
|---|---|
| 0.1 | 1 in 10 (E-1) |
| 0.01 | 1 in 100 (E-2) |
| 0.001 | 1 in 1000 (E-3) |
| 0.0001 | 1 in 10,000 (E-4) |
| 0.00001 | 1 in100,000 (E-5) |
| 0.000001 | 1 in 1,000,000 (E-6) |

**Table 3**

| Threat | Progression (failure mode) | Likelihood of failure for independent FM | Considered for overall probability | |
|---|---|---|---|---|
| Floods | Crest overtopping | 5E-6 | 5E-5 | Take highest for each external threat |
| | Chute overtopping | 5E-5 | | |
| Internal Threats | Body of dam | 6E-4 | 6E-4 | Include all failure modes for internal threats |
| | Foundation | 6E-6 | 6E-6 | |
| | Interface between structure and embankment | 6E-5 | 6E-5 | |
| Overall likelihood of failure | | 1.8E-4 | 1.8E-4 | |

**Table 4**

| Station | T | Frequency | Station | T | Frequency |
|---|---|---|---|---|---|
| Lerwick | 102 | 0.0098 | Wick | 102 | 0.0098 |
| Aviemore | 8 | 0.125 | Abbotsinch | 10,000 | 0.0001 |
| Leeming/Dishforth | 890 | 0.0011 | Manchester | 10,000 | 0.0001 |
| Elmdon | 10,000 | 0.0001 | Honington/Mildenhall | 10,000 | 0.0001 |
| Doscombe Down | 7,900 | 0.0012 | Plymouth | 10,000 | 0.0001 |
| Valley | 10,000 | 0.0001 | Aldergrove | 10,000 | 0.0001 |
| Redesdale | 84 | 0.0119 | Moor House | 4 | 0.25 |
| Malham Tarn | 4 | 0.25 | Wilsden | 29 | 0.0345 |
| Cwmystwyth | 80 | 0.0125 | Widdybank Fell | 2 | 0.5 |

**Table 5**

| |
|---|
| **Physical processes incorporated into WRF**. |
| * Generation of PBL turbulence and related transports (including non-local effects but essentially dry surface-based processes.) |
| * Surface-atmosphere exchanges momentum, heat, moisture, might eventually include other quantities, either land or water surfaces) including dependencies on surface and subsurface processes. |
| * Generation of subgrid-scale turbulence and related transports above the PBL (resulting in primarily local diffusion) |
| *Convection (non-local fluxes aided by condensation, including "shallow" convection) |
| * Radiation (short and long wave, atmospheric and surface effects) |
| * Cloud and precipitation physics (local processes including fallout) |

**Table 6**

| MP Physics | Scheme |
|---|---|
| 8 (/98) | Thompson (old) |
| 9 | Milbrandt 2 - moments |
| 10 | Morrison 2 - moments |
| 16 | WDM6 |

**Figure legends**

**Figure 1:** (a) Distribution of Category A (red) and B (yellow) dams; (b) an example of the area likely to be flooded should dams fail near Pontefract, Yorkshire; note the route of the M62 motorway running west to east and the A1(T) running north to south (courtesy Environment Agency.

**Figure 2:** Rainfall 19$^{th}$ June 2005 in the area of the Boltby reservoir, 125 mm in 3 hours. Also shown is the orography (from Wass et al., 2008).

**Figure 3:** Accumulated WRF rainfall 12-18Z 19 June 2005 using (a) Lin, (b) Thompson or Millbrandt, (c) Morrison, (d) WDM6 microphysics schemes and (e) certain "features" have been added so that comparison can be made with Fig. 2. Topography is colour shaded in metres. Rainfall contour interval in (a) to (d) is 10mm.

**Figure 1**

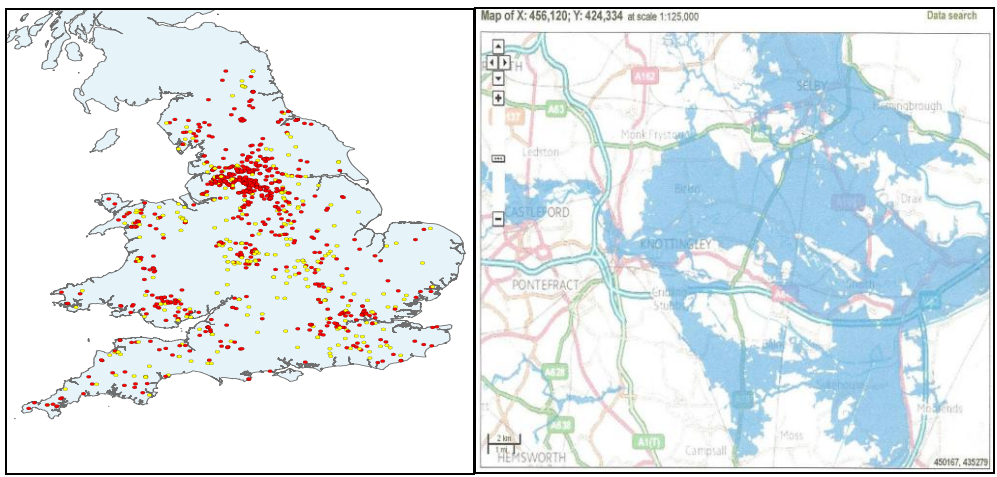

      **(a)**                                 **(b)**

**Figure 2**

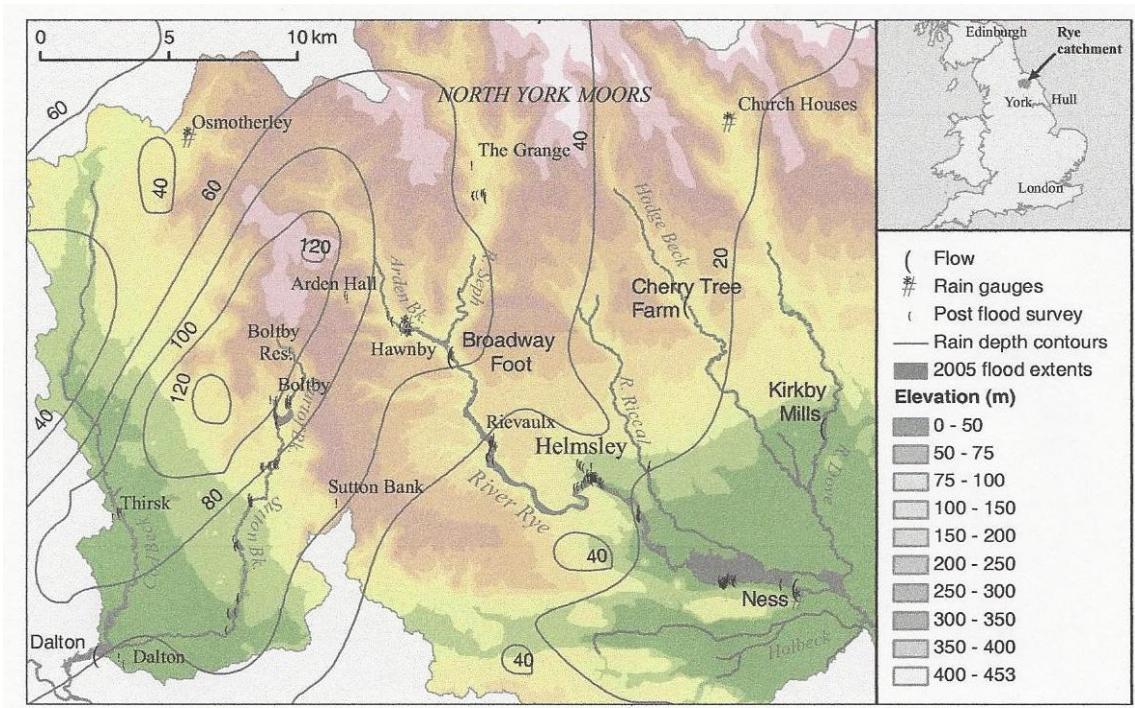

**Figure 3**

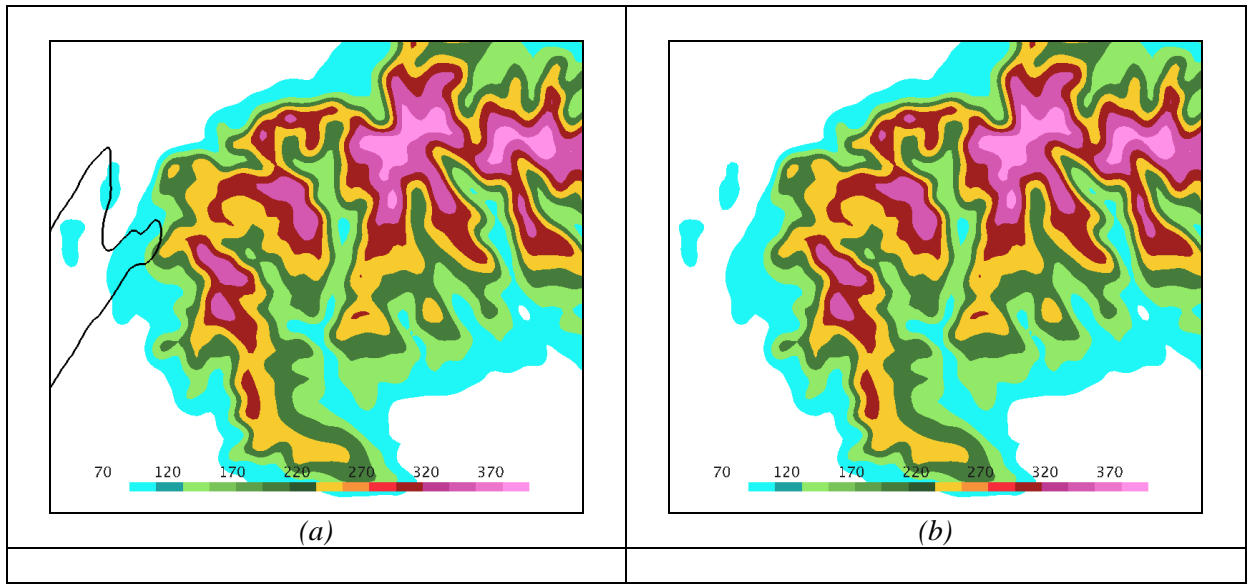

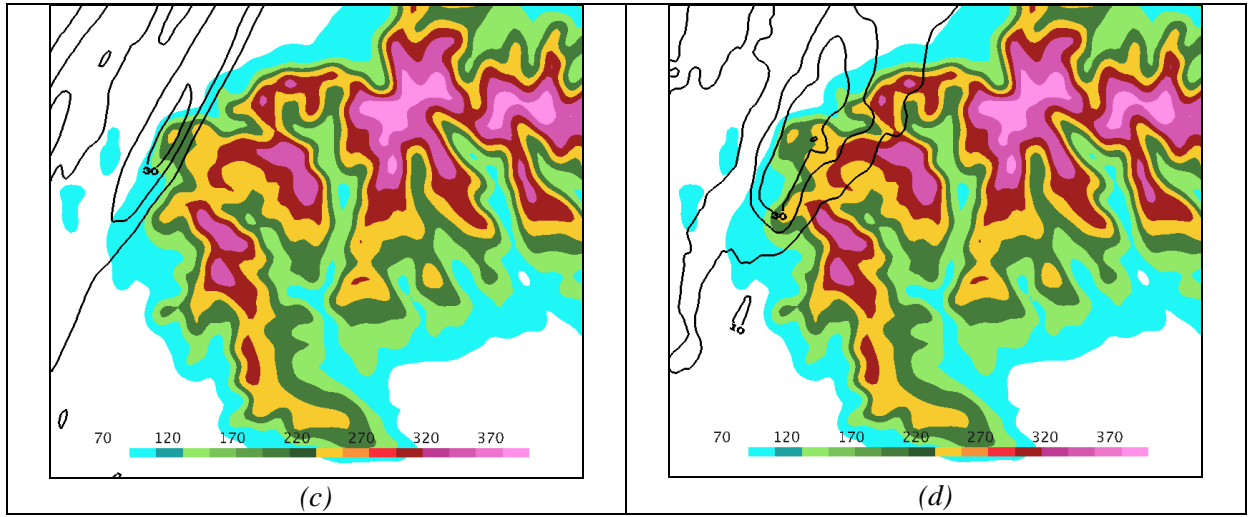

*(c)*                                         *(d)*

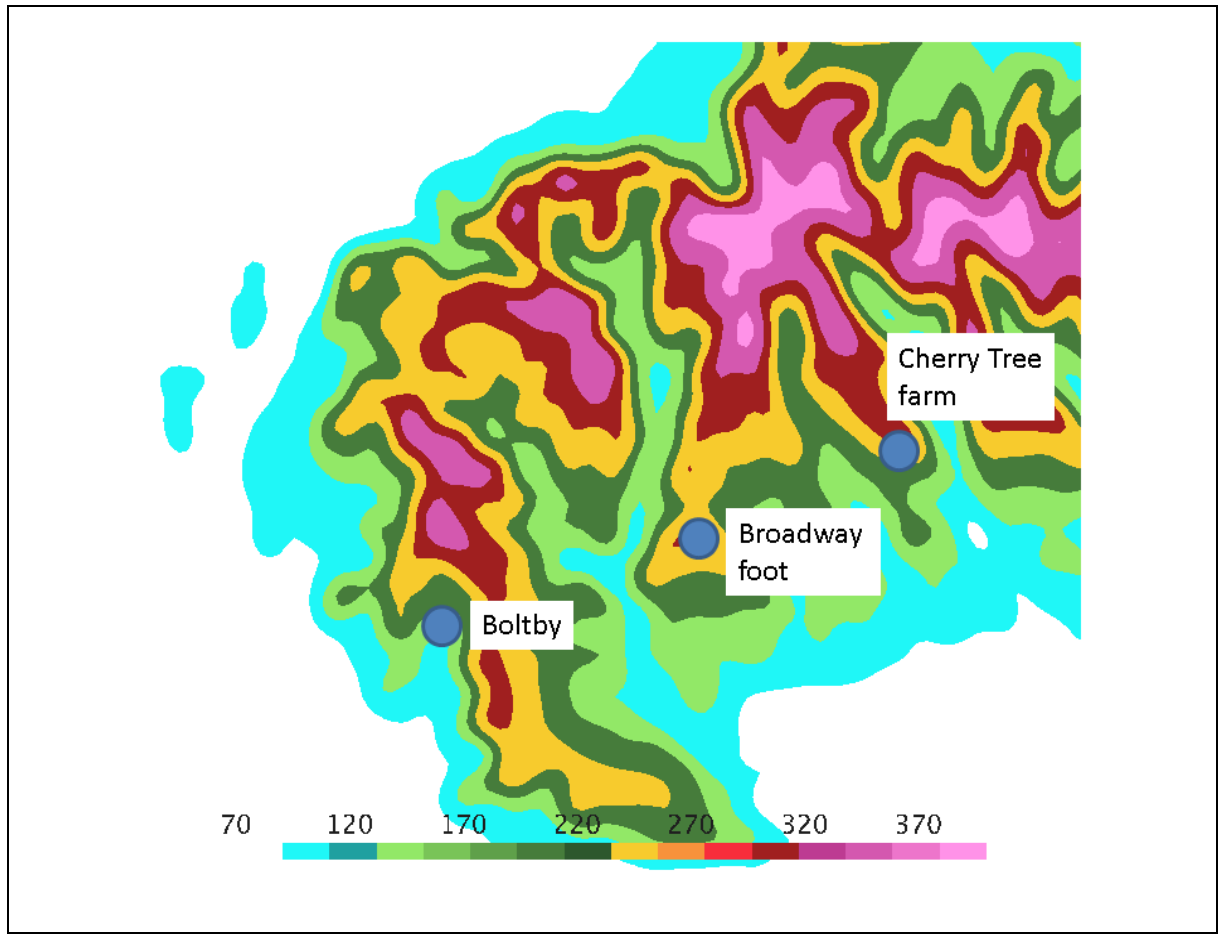

*(e)*